\documentclass[11pt,a4paper]{article}

\usepackage[a4paper,
  top=25mm,
  bottom=30mm,
  left=20mm,
  right=20mm,
  columnsep=7mm
]{geometry}

\usepackage{siunitx}
\usepackage[dvipdfmx]{graphicx}
\usepackage{amsmath,amssymb,bm}

\usepackage{cite}

\usepackage{url}

\usepackage{xcolor}
\colorlet{blue}{black}

\title{Machine Learning of Temperature-dependent Chemical Kinetics Using Parallel Droplet Microreactors}

\author{Mamoru Saita, \ Yutaka Hori\thanks{Corresponding author: Yutaka Hori, e-mail: \tt{yhori@keio.jp}} \bigskip \\ 
Department of Applied Physics and Physico-Informatics, \\Faculty of Science and Technology,
Keio University, Kanagawa, Japan
}
\date{}
\begin{document}

\maketitle

\section*{Abstract}
Temperature is a fundamental regulator of chemical and biochemical kinetics, yet capturing nonlinear thermal effects directly from experimental data remains a major challenge due to limited throughput and model flexibility. Recent advances in machine learning have enabled flexible modeling beyond conventional physical laws, but most existing strategies remain confined to surrogate models of end-point yields rather than transient dynamics. Consequently, an end-to-end framework unifying systematic kinetic data acquisition with machine learning based modeling has been lacking. In this paper, we present a unified framework that integrates droplet microfluidics with machine learning for the systematic analysis of temperature-dependent reaction kinetics. The platform is designed to enable immobilization and long-term time-lapse imaging of thousands of droplets under dynamic thermal gradients. This configuration yields massively parallel time-resolved datasets across diverse temperature conditions capturing transient kinetics and provides suitable inputs for training machine-learning models of reaction dynamics. Leveraging these datasets, we train Neural ODE models {\color{blue}to learn nonlinear temperature-dependent kinetics directly from experimental data}. We demonstrate accurate prediction of enzymatic kinetics, highlighting the robustness and versatility of the approach. Our framework bridges high-throughput experimental data acquisition with data-driven modeling, establishing a foundation for predictive ability and rational analysis and design of temperature-sensitive biochemical processes.

\vspace{0.5cm}

\noindent
\textbf{Keywords:} Droplet microfluidics, Neural ODE, Data-driven modeling, Temperature-dependent kinetics

\section{Introduction}
Reactions in chemistry and biology are highly sensitive to their surrounding conditions. 
Among many parameters, temperature is one of the central regulators that globally governs chemical and biological kinetics \cite{Gillooly2001,Verghese2012,Moon2023} and has been studied extensively across diverse contexts from fundamental chemical processes \cite{Somero1995,Richter2010,Bisswanger2014,Robinson2015} to engineered systems in synthetic biology \cite{Kortmann2012,Chee2022,Choi2022,Laohakunakorn2020}. 
{\color{blue}A major challenge in understanding and predicting temperature-dependent reaction kinetics is that reaction rates generally exhibit a nonlinear dependence on temperature, such that even small temperature variations can result in substantial changes in reaction dynamics.}
Consequently, elucidating how reaction dynamics depend on thermal conditions has been a longstanding objective, motivating many experimental and theoretical efforts.

While conventional kinetic models incorporate temperature through physical laws \cite{Daniel2010,Arcus2020} such as the Arrhenius equation \cite{Laidler1984}, capturing {\color{blue}these nonlinear temperature-dependent kinetics} with high accuracy from experimental data remains challenging due to unmodeled dynamics and uncertainties inherent in chemical systems. Recent advances in machine learning, however, have enabled the development of more flexible models with greater expressive capacity. These machine learning models enhance predictive ability and provide deeper insight into reaction dynamics when 
combined with mechanistic models.
To date, most learning based approaches optimize the end-point yields of reactions by constructing surrogate models, for example, through Bayesian optimization \cite{Shields2021,Clayton2020,Jeraal2021,Taylor2023brief} or neural networks \cite{Granda2018,Zhu2025}.{\color{blue}
While such endpoint models are useful for reaction optimization, they provide limited information about the temporal evolution of reactions or their responses to dynamically changing environmental conditions. 
Predictive kinetic models are therefore particularly valuable for analyzing biochemical systems operated under dynamically varying temperature conditions, as well as larger reaction networks in which the timing of upstream reactions influences downstream dynamics.
In this context}, Neural ODEs \cite{Chen2018} have recently been demonstrated as an effective formalism for modeling the transient dynamics of reaction kinetics \cite{Thoni2025,Beheler-Amass2025}, which are particularly critical in systems where time-dependent responses shape overall behavior \cite{Elowitz2000,Niederholtmeyer2015,Hsiao2016,Aoki2019,deCesare2022,Chatani2023}. Yet, these data-driven models are often constrained by experimental throughput, highlighting the need for an end-to-end framework that unifies large-scale kinetic data generation with machine learning while minimizing sample consumption.

Droplet microfluidics offers a promising approach for massively parallel measurements of reaction dynamics under diverse conditions with minimal sample consumption. 
This capability has previously been exploited to screen parameters and support modeling and analysis of synthetic biomolecular systems such as genetic networks and DNA circuits \cite{Genot2016,Hori2017,Baccouche2017}. Temperature dependence has also been explored by generating thermal gradients across droplet chambers \cite{Laval2007,Lobato2023}. Despite their capacity to generate large-scale datasets, existing implementations remain disconnected from machine learning frameworks capable of capturing reaction kinetics, leaving a gap between data acquisition and data-driven understanding of temperature-dependent kinetics.

In this paper, we address this gap by integrating droplet microfluidics with machine learning to systematically analyze temperature-dependent kinetics of chemical reactions. The proposed framework combines massively parallel droplet assays with thermal control, enabling the acquisition of large-scale and temperature-resolved datasets for machine learning. 
At the core of this framework is a droplet chamber that suppress thermal convection and allows long-term immobilization of droplets under temperature gradients.
This design permits fluorescence based kinetic measurements immediately after droplet generation, yielding datasets that capture early transient dynamics essential for data-driven kinetic modeling. 
Leveraging these datasets, we construct neural-network based models capable of predicting enzymatic kinetics across diverse thermal environments. In particular, we adopt Neural ODEs \cite{Chen2018}, which embed neural networks within the structure of differential equations with temperature as an input parameter. 
{\color{blue}Our Neural ODE formulation retains the known reaction stoichiometry while representing only the unknown temperature-dependent reaction-rate function with a neural network, thereby combining mechanistic interpretability with the flexibility of data-driven modeling, in contrast to purely sequence-based architectures for time-series modeling.
This formulation enables the temperature-dependent reaction rate to be learned directly from experimental time-series data without assuming a predefined rate law.
Furthermore, the learned reaction-rate function remains compatible with conventional ODE based reaction models and can be extended to more complex reaction networks.}
The proposed end-to-end framework bridges high-throughput experimental data acquisition with data-driven modeling and provides a versatile platform for prediction and design of biochemical processes under controlled thermal environments.

\section{Results}
\subsection{Concept of Framework and Workflow}
\begin{figure*}
 \centering
 \includegraphics[width=\textwidth]{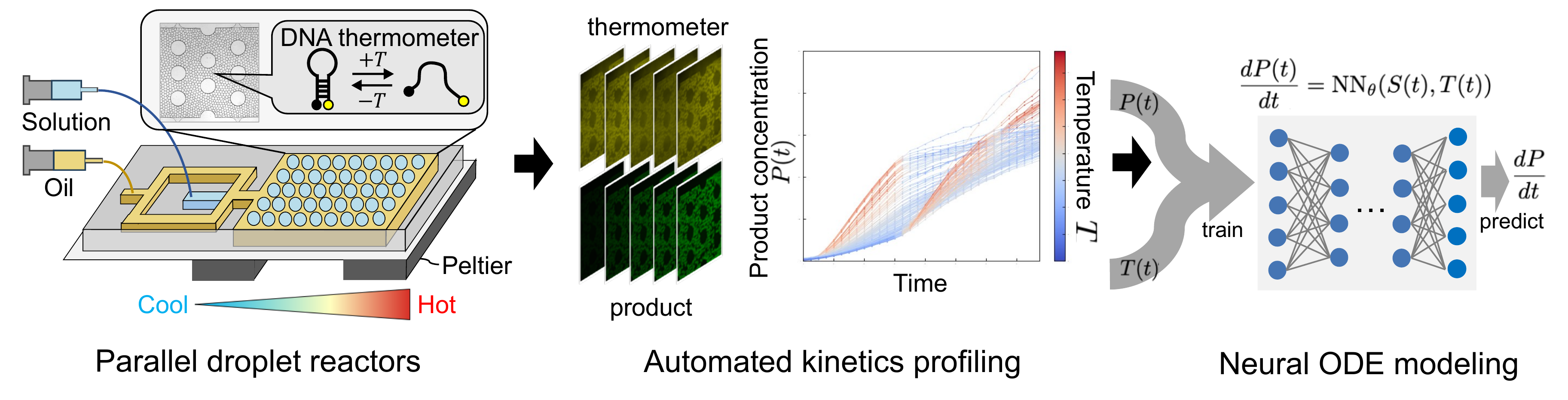}
 \caption{
 \textbf{End-to-end framework for data-driven modeling of temperature-dependent kinetics.}\\
A microfluidic device generates and immobilizes thousands of droplets under thermal gradients with droplet temperatures monitored by DNA thermometers. Automated microscopy and image analysis enable tracking of time-resolved fluorescence signals, capturing long-term kinetics across diverse thermal conditions. The resulting large-scale time-series datasets are used to train neural networks embedded in ordinary differential equations (Neural ODEs), which flexibly capture nonlinear temperature dependencies.}

 \label{fgr:fig1_concept}
\end{figure*}

We present an integrated framework that links high-throughput droplet microfluidics with machine-learning based modeling of temperature-dependent chemical kinetics. 
The proposed workflow consists of (i) droplet based microfluidics for generating and immobilizing thousands of parallel reactors under controlled temperature gradient, (ii) automated in situ imaging and tracking of both droplet temperature and product concentration, and (iii) data-driven modeling of temperature-dependent reaction kinetics using Neural ODE, a neural network architecture embedded in differential equations (Fig. \ref{fgr:fig1_concept}). 
In the microfluidic platform, water-in-oil emulsion droplets containing reaction mixtures and a DNA based fluorescence thermometer \cite{Jonstrup2013, Gareau2016} are generated and immobilized in {\color{blue}the same} observation chamber. 
{\color{blue}The integration of droplet generation, immobilization, and measurement within a single device enables time-lapse imaging to begin immediately after enzyme-substrate mixing, allowing early-stage reaction dynamics to be captured for kinetic modeling.}
Temperature gradients are applied to droplets using a custom Peltier-based platform placed under the observation chamber. 
By tracking time-lapse fluorescence signals of the DNA based temperature probe and the reaction output, the platform enables massively parallel characterization of reaction kinetics under diverse thermal conditions.
The resulting kinetic profiles are then used to train the Neural ODE model. 
In particular, the use of Neural ODE enhances the flexibility of the model beyond conventional ODE formulations, allowing the nonlinear temperature effects to be captured directly from experimental data  thereby bridging the gap between data acquisition and predictive modeling.

\subsection{Parallel droplet reactors under controlled temperature gradients}
\begin{figure*}
 \centering
 \includegraphics[width=\textwidth]{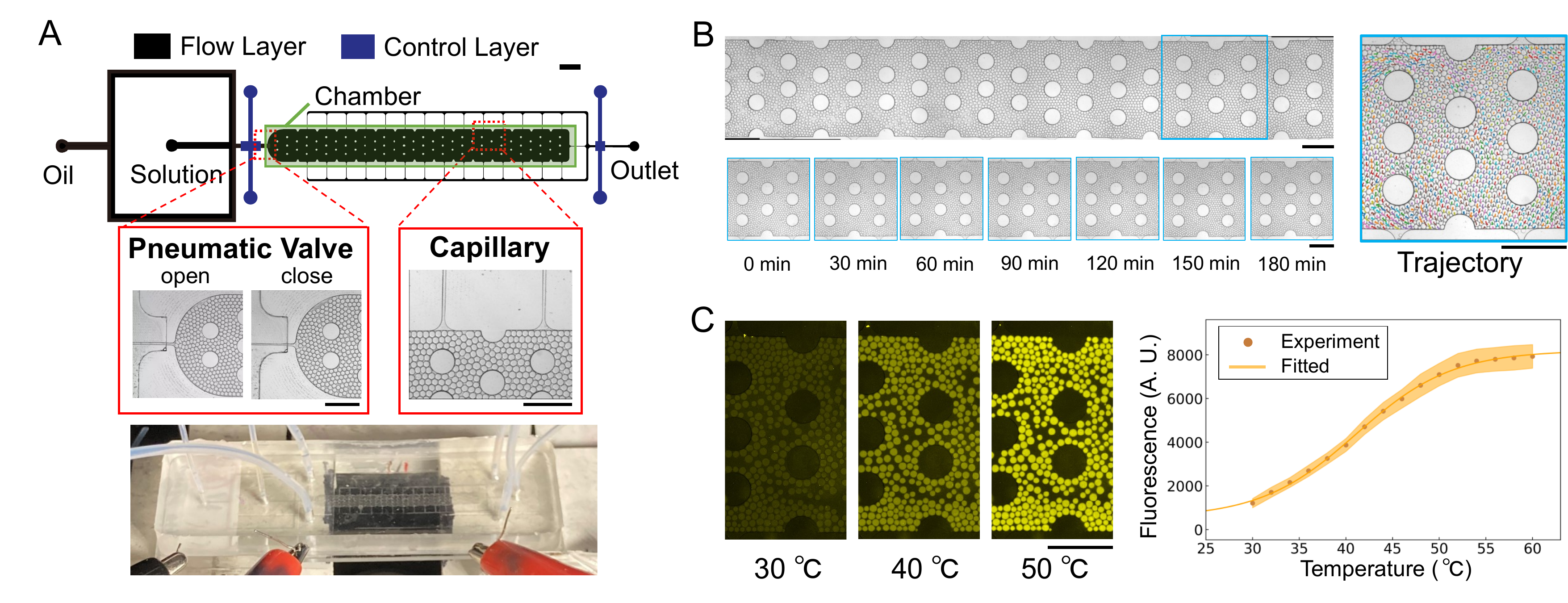}
 \caption{
 \textbf{Microfluidic platform for droplet immobilization and temperature readout}.\\
(A) Schematic of the two-layer microfluidic device integrating droplet generation and immobilization under thermal gradients. 
(B) Experimental validation showing stable confinement of droplets over extended periods without coalescence or displacement. Trajectory analysis of droplet centers over 180 min confirms that droplets remain immobilized with minimal displacement.
(C) Fluorescence images of droplets encapsulating DNA thermometers under stationary temperature gradients. Calibration of fluorescence intensity to temperature. 
{\color{blue}The larger circular features visible in panels B and C correspond to the PDMS pillar structures of the observation chamber.}
Scale bars represent 1 mm.
}
 \label{fgr:fig2_verification}
\end{figure*}

To realize this workflow, we designed a two-layer PDMS microfluidic device that integrates droplet generation with high-density immobilization in an observation chamber (Fig. \ref{fgr:fig2_verification}A). 
The capillary structure in the observation chamber removes the oil phase and facilitates dense confinement of droplets, and pneumatically actuated valves seal the chamber to suppress fluid exchange. These features minimize the influence of thermal convection and allow reliable long-term time-series observation under thermal gradient.

To demonstrate the capability of long-term data acquisition, we first examined droplet immobilization and tracking under thermal gradients. 
Specifically, droplets were immobilized in the observation chamber and subjected to a temperature gradient ranging from \SI{30}{\celsius} to \SI{60}{\celsius}. The droplets remained compartmentalized without coalescence for at least 180 min (Fig. \ref{fgr:fig2_verification}B, Supplementary Note~1, {\color{blue}Fig. S1}). Moreover, the time-lapse bright-field images revealed that the droplet displacement was limited to approximately one droplet radius (100 {\textmu}m), ensuring that the same droplets could be reliably tracked using a custom computer program to obtain time-series data for modeling.

Next, we assessed the ability to read out the temperature of individual droplets. Droplets encapsulating DNA based fluorescent thermometers were immobilized in the observation chamber and exposed to a spatially uniform temperature field. 
A graded fluorescence intensity was observed across the temperature range of \SI{30}{\celsius} to \SI{60}{\celsius}, yielding a calibration curve relating fluorescence intensity to temperature (Fig.~\ref{fgr:fig2_verification}C, Supplementary Note~2, {\color{blue}Table S1}). 
{\color{blue}The precision of temperature readout 
and the temporal response of the DNA thermometer were further characterized as described in Supplementary Note~2 and Fig.~S2.} 
This response was consistent with the temperature-dependent melting of DNA duplexes, confirming that temperature of individual droplets can be measured in situ along with the reaction output.

These results demonstrate that the framework enables both stable and long-term immobilization of droplets and acquisition of reliable temperature dependent output of reactions at the single-droplet level, establishing the basis for subsequent kinetics measurements and systematic data-driven modeling. 

\subsection{Application to enzymatic kinetics under stationary temperature gradients}
\label{sec:stationary}
\begin{figure*}[p]
 \centering
 \includegraphics[width=0.92\textwidth]{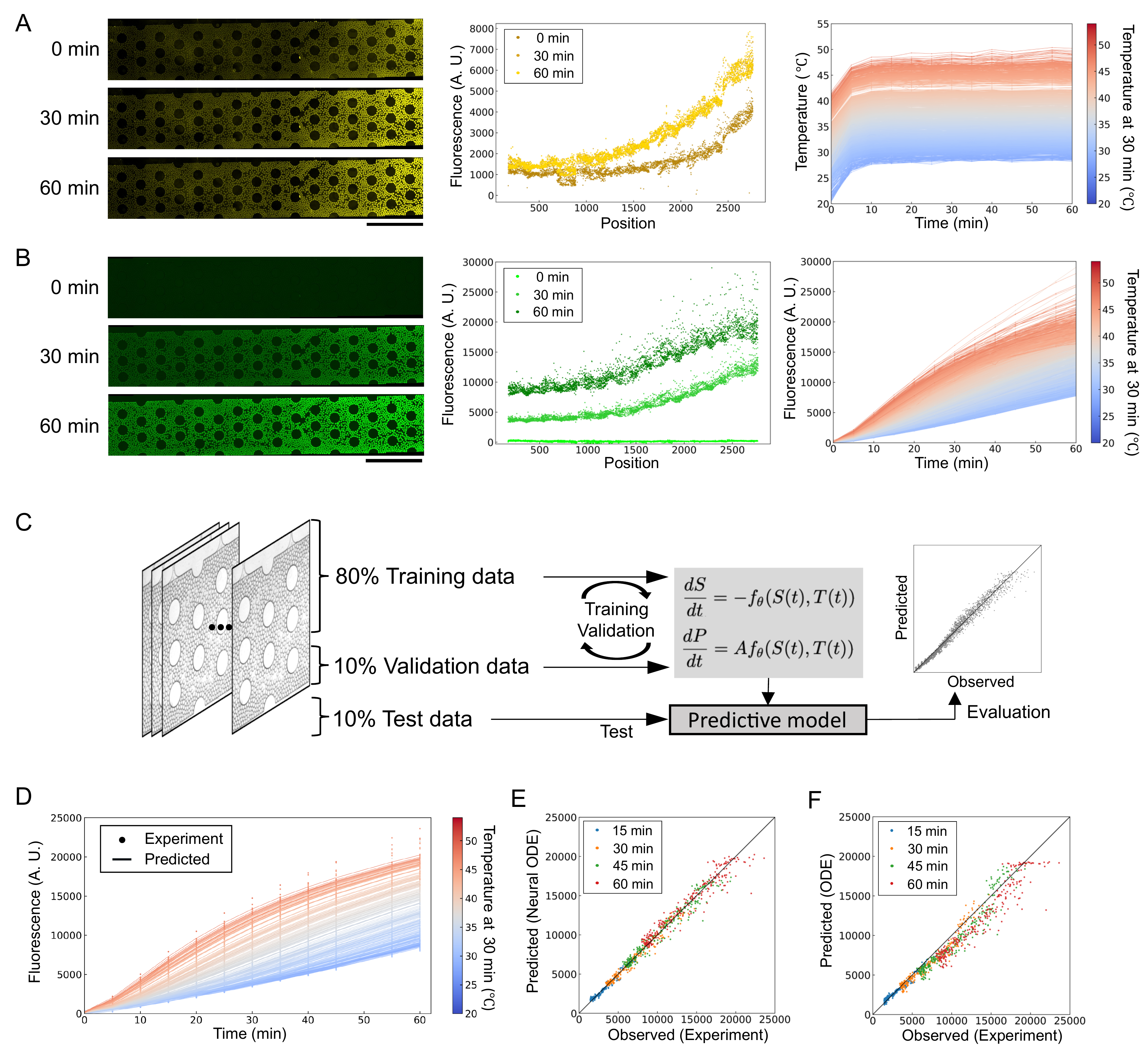}
 \caption{
 \textbf{Parallel measurements of temperature-dependent enzymatic kinetics and Neural ODE based modeling}.\\
(A) Fluorescence images of DNA thermometers in droplets under a constant temperature gradient. 
Left: fluorescence images over time. Center: fluorescence intensity profiles of droplets along the chamber position. 
Right: time-series of droplet temperatures obtained using the calibration curve in Fig. \ref{fgr:fig2_verification}C.
(B) Fluorescence images of enzymatic reaction products under a constant temperature gradient. 
Left: fluorescence images of droplets over time. 
Center: fluorescence intensity profiles of droplets along the chamber position. 
Right: time-series of product fluorescence. 
(C) Workflow of Neural ODE modeling. Time-series data were split into 80\% training, 10\% validation, and 10\% test sets. Hyperparameters were tuned using validation data, and the final model was evaluated using test data. 
(D) Time-series of product fluorescence in the test dataset, corresponding to the right panel of B, together with predictions by the Neural ODE model.
(E) Predicted fluorescence obtained by Neural ODE plotted against experimental data. 
(F) Predicted fluorescence obtained by an ODE model with Arrhenius rate law plotted against experimental data, showing the advantage of the proposed framework in capturing nonlinear temperature dependencies. 
Scale bars represent 3 mm.
}
 \label{fgr:fig3_constant_gradient}
\end{figure*}

To showcase the capability of the proposed framework, we next performed measurements of 
enzymatic reactions under a stationary temperature gradient.
A a model system, we used $\beta$-glucosidase, a hydrolytic enzyme that specifically cleaves $\beta$-D-glycosidic bonds, and fluorescein-di-$\beta$-D-glucopyranoside (FDGlu) as the substrate, which releases fluorescein upon enzymatic cleavage. 
{\color{blue}This relatively simple and well-characterized enzymatic reaction serves as a suitable model system for validating the proposed framework without complications arising from complex reaction pathways or poorly understood mechanisms.
The reaction can be described by the following enzymatic reaction scheme}
\begin{align}
S + E \leftrightharpoons S\!:\!E \rightarrow P,
\label{eq:reaction_scheme}
\end{align}
where $S$, $E$, and $S\!:\!E$ represent FDGlu (substrate), $\beta$-glucosidase (enzyme), and their substrate-enzyme complex, respectively, and $P$ denotes the fluorescent product generated by the cleavage reaction. 
Since $\beta$-glucosidase does not cleave the N-glycosidic bonds present in the DNA thermometer, the DNA thermometer remains intact during the measurement. 
{\color{blue}In addition, since the FDGlu concentration was in the micromolar range, the thermal effect of the enzymatic reaction was negligible compared with the imposed temperature gradient.}

Droplets containing the enzymatic reaction components and the DNA thermometer were generated {\color{blue}on chip} and immediately immobilized in {\color{blue}same} the observation chamber preconditioned with a stationary temperature gradient. 
To quantify the temperature dependent reaction kinetics, we developed a custom image processing program that tracked the fluorescence intensity of individual droplets and generated time-series data for more than 2000 droplets. 
As shown in Figs. \ref{fgr:fig3_constant_gradient}A and \ref{fgr:fig3_constant_gradient}B, droplets located at higher temperatures exhibited faster increases in product fluorescence, consistent with the expected thermal dependence of enzymatic activity (see also Supplementary Movies 1 and 2). 
The temperature of each droplet was estimated from the fluorescence of the co-encapsulated DNA thermometer using the calibration curve (Fig. \ref{fgr:fig2_verification}C). 
{\color{blue}Importantly, the in situ measurement enables the actual temperature of each droplet to be directly incorporated into the subsequent kinetic modeling.
Moreover}, the integration of droplet generation and measurement within a single device enabled fluorescence imaging to begin within minutes of enzyme-substrate mixing. This temporal advantage is critical particularly for capturing the early-stage dynamics of the enzymatic reaction across a range of temperatures, providing high-resolution time-series data that are well suited for training machine learning models of temperature-dependent kinetics.

We leveraged these datasets to construct and train Neural ODE models, in which the reaction rate is represented by a neural network embedded in ordinary differential equations. 
{\color{blue}Specifically, based on the enzymatic reaction scheme in eq.~(\ref{eq:reaction_scheme}), we represent only the unknown temperature-dependent reaction-rate function using a neural network. The resulting kinetic model is given by}
\begin{align}
& \frac{dS(t)}{dt} = -f_\theta(S(t), T(t)), \label{eq:S}\\
&  \frac{dP(t)}{dt} = A f_\theta(S(t), T(t)) \label{eq:P}, 
\end{align}
where $P(t)$ is the observed fluorescence signal corresponding to the concentration of the product, $S(t)$ is a latent variable corresponding to {\color{blue}the spatially averaged substrate concentration within a droplet}, $T(t)$ is the temperature, and $A$ is a constant. 
{\color{blue}Eq. \eqref{eq:S} describes substrate consumption, while eq. \eqref{eq:P} describes product formation through the same reaction-rate function, thereby preserving the coupling between substrate consumption and product formation implied by the reaction scheme.}
Instead of prescribing $f_\theta(S(t), T(t))$ with a fixed functional form, we here used a neural network to parameterize the function $f_\theta(\cdot)$ with $\theta$ being the parameters of the network to be trained. This enables the model to capture nonlinear temperature dependencies that are difficult to represent with conventional physical rate laws.
{\color{blue}Moreover, once trained, the model can predict reaction trajectories under arbitrary temperature profiles, including conditions that were not directly measured experimentally. Such predictive kinetic models are particularly useful for analyzing biochemical systems operated under dynamically varying thermal environments.}
{\color{blue}It should be noted that each droplet is assumed to behave as a well-mixed microreactor due to the relatively fast intradroplet diffusion compared with the reaction timescale and the large number of molecules per droplet.}

To train and evaluate the model, we first preprocessed the dataset by removing outliers {\color{blue}from imaging artifacts. Specifically, droplets located within the outermost 5\% of the observation chamber on both lateral sides were excluded from the analysis since fluorescence measurements in these regions were consistently lower than those in the central region, likely due to shading from the Peltier elements (Fig. S4). 
This preprocessing resulted in } approximately 2000 time-series datasets. 
These datasets were then randomly partitioned into training, validation, and test sets at a ratio of 8:1:1 for each temperature bin (Fig.~\ref{fgr:fig3_constant_gradient}C).
The model was trained using the training dataset by minimizing a loss function defined as the weighted mean-squared error between simulated and measured fluorescence signals (see Section~\ref{sec:experimental} for details). The validation dataset was used for hyperparameter tuning during the training.

Finally, the predictive performance of the trained Neural ODE model was evaluated using the test dataset. As shown in Fig.~\ref{fgr:fig3_constant_gradient}D, the model predictions closely matched the experimentally observed fluorescence trajectories across the entire temperature range.
In Fig.~\ref{fgr:fig3_constant_gradient}E, the predictive accuracy is further quantified using a scatter plot comparing predicted and experimentally observed fluorescence values at $t = 15, 30, 45$, and $60$ min in the test dataset. The data points closely align with the identity line, indicating strong agreement between predicted and observed values. The coefficient of determination computed over all data points is {\color{blue}$R^2 = 0.987$} (Supplementary Note~3, {\color{blue}Fig. S3}).
{\color{blue}To evaluate the robustness of the selected neural network architecture, we also performed an architecture sensitivity analysis by varying the number of hidden layers and the number of neurons per hidden layer (Supplementary Note 3). Once sufficient model capacity was provided, the predictive performance remained nearly unchanged across the tested architectures, indicating that the learned kinetics are not sensitive to the exact network architecture (Table S2). Similarly, varying the lateral-edge exclusion width had little effect on the predictive performance, indicating that the learned model is robust to the preprocessing of outliers (Table S3).}

To benchmark the Neural ODE model against a conventional physics based approach, we further constructed a mechanistic ODE model based on mass-action kinetics with an Arrhenius-type temperature dependence. The model parameters were then fitted to the experimental data (Supplementary Note~4, Figs. S5 and S6). The best-fit conventional ODE model reproduced the overall trend of the experimental data across the temperature range (Fig.~S6B). However, quantitative comparison revealed systematic underestimation of the reaction dynamics relative to the experimental observations as shown in Fig.~\ref{fgr:fig3_constant_gradient}F. The coefficient of determination computed over all data points is {\color{blue}$R^2 = 0.956$} (Fig.~S6C).
This discrepancy suggests that the simple Arrhenius formulation is insufficient to fully capture the temperature-dependent reaction dynamics. 

These results demonstrate the capability of our platform to rapidly generate large and time-resolved datasets across diverse thermal conditions and to harness them for training machine-learning models of reaction kinetics with high predictive accuracy, highlighting its potential as a versatile tool for data-driven exploration of chemical kinetics.

\subsection{Application to enzymatic kinetics under dynamic temperature gradients}
\begin{figure*}[tb]
 \centering
 \includegraphics[width=\textwidth]{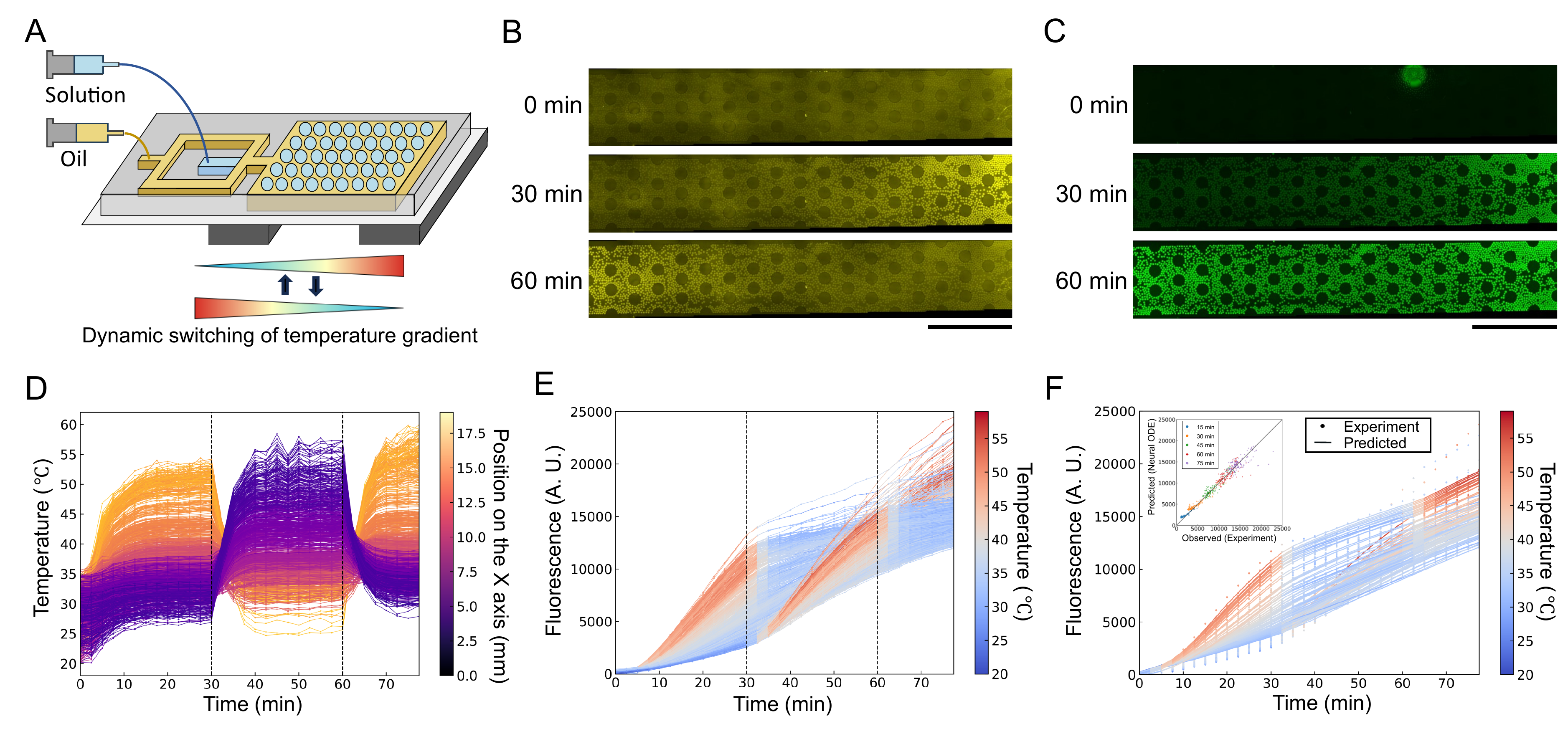} 
 \caption{ \textbf{Temperature-dependent kinetics under dynamic temperature gradients and Neural ODE based modeling}.\\
(A) Schematic of the experimental setup showing a dynamic temperature gradient with switching at $t=30$ and $t=60$ min.
(B) Fluorescence images of DNA thermometers in droplets under a dynamic temperature gradient. 
(C) Fluorescence images of enzymatic reaction products in droplets under a dynamic temperature gradient. 
(D) Time-series of droplet temperatures obtained from DNA thermometer fluorescence (panel B). 
(E) Time-series of product fluorescence across diverse temperature range (panel C). 
(F) Time-series of product fluorescence compared with the prediction by Neural ODE model. The inset shows a scatter plot comparing Neural ODE predictions with experimental data, showing close agreement.
Scale bars represent 3 mm.}
 \label{fgr:fig4_change_gradient}
\end{figure*}

Finally, we evaluated the capability of the proposed framework to perform high-throughput measurements of enzymatic reaction kinetics under dynamically varying temperature conditions. To this end, droplets containing both the enzymatic reaction mixture and the DNA thermometer were generated following the same protocol described in Section~\ref{sec:stationary}.
To impose a time-varying temperature gradient, the temperatures of the two Peltier elements located beneath the left and right sides of the microfluidic device were initially set to \SI{30}{\celsius} and \SI{65}{\celsius}, respectively, and subsequently swapped at $t = 30$~min and $t = 60$~min after the start of the experiment. This operation established a dynamically varying temperature field across the observation chamber (Fig.~\ref{fgr:fig4_change_gradient}A).

Using this setup, we acquired time-lapse images of both the DNA thermometer and the enzymatic reaction output from more than 2000 droplets (Figs.~\ref{fgr:fig4_change_gradient}B and \ref{fgr:fig4_change_gradient}C, Supplementary Movies 3 and 4). 
The fluorescence signal of the DNA thermometer dynamically responded to the temporal changes in the temperature field, confirming its function as an in situ temperature probe (Fig.~\ref{fgr:fig4_change_gradient}D). Moreover, the fluorescence intensity of the reaction product within individual droplets increased more rapidly during high-temperature intervals and slowed during lower-temperature phases, consistent with the expected temperature dependence of enzymatic activity. 
Based on these measurements, time-resolved training datasets capturing the enzymatic reaction kinetics across dynamically varying temperatures were obtained (Fig.~\ref{fgr:fig4_change_gradient}E).

These datasets were partitioned into training, validation, and test sets following the same protocol as in Section~\ref{sec:stationary} (Fig.~\ref{fgr:fig3_constant_gradient}C), and the Neural ODE model was trained. 

The trained model was subsequently validated using the test dataset. 
As illustrated in Fig. \ref{fgr:fig4_change_gradient}F, the trained model successfully reproduced the experimental time-series under dynamic temperature shifts (see also Supplementary Note 5, {\color{blue}Fig. S7}). 
{\color{blue}Similar to the stationary temperature-gradient experiment, we performed an architecture sensitivity analysis by varying the number of hidden layers and the number of neurons per hidden layer. Although the dynamic temperature-gradient dataset required a larger model capacity than the stationary case, architectures with sufficient model capacity achieved comparable predictive performance, indicating that the learned kinetics are robust to the exact network architecture (Table S4). The model robustness was also confirmed to the choice of lateral-edge exclusion width (Table S5, Fig. S8).}

These results demonstrate that the proposed framework enables efficient learning of reaction kinetics even for time-varying thermal conditions.

\section{Conclusion}
We have developed an end-to-end framework that unites droplet microfluidics with machine learning based modeling to enable systematic and high-throughput exploration of temperature-dependent chemical kinetics. The proposed framework integrates (i) droplet generation and immobilization under controlled thermal gradients, (ii) automated {in situ} imaging of droplet temperature and product concentration across thousands of droplets, and (iii) Neural ODE based modeling of kinetics, forming a unified workflow for data-driven analysis of temperature-dependent chemical reactions. This capability was validated by experiments demonstrating stable droplet immobilization and {in situ} temperature readout by DNA thermometers (Fig.~\ref{fgr:fig2_verification}), establishing the basis for large-scale data acquisition under dynamic thermal gradients. Building on this foundation, we acquired thousands of time-series datasets that simultaneously captured droplet temperature and reaction output. These datasets enabled the training of Neural ODE models, which flexibly represented nonlinear temperature dependencies beyond conventional formulations (Fig.~\ref{fgr:fig3_constant_gradient}). Validation across both stationary and dynamic temperature gradients confirmed that the trained models robustly reproduced biochemical kinetics under diverse conditions (Figs.~\ref{fgr:fig3_constant_gradient} and \ref{fgr:fig4_change_gradient}). These results demonstrate how the proposed framework transforms high-throughput experimental measurements into highly expressive data-driven models and bridges the gap between large-scale experimentation and computational modeling.

The proposed framework is especially suited to applications where transient kinetics govern functional performance  such as biomolecular circuits in synthetic biology and non-equilibrium chemistry. 
The framework provides a powerful means to characterize the dynamics of reaction systems and facilitate the rational design of thermally robust and responsive systems through automated high-throughput experimentation.
Recent advances in machine learning models have shown strong potential for representing complex nonlinear kinetics, but their flexibility often comes at the expense of interpretability.
Embedding physics-informed structures provides a promising route to address this challenge by combining the flexibility of machine learning model with mechanistic insight \cite{Fedorov2023}. Integrating such models with the proposed platform is expected to further enhance predictive design and analysis of biochemical networks operating under dynamic thermal environments, opening new opportunities for uncovering design principles of nonequilibrium chemical and biological processes.

\section{Experimental Section}
\label{sec:experimental}
\subsection*{Device fabrication} 
The microfluidic device was fabricated by multi-layer soft lithography using polydimethylsiloxane (PDMS; SILPOT 184, Dow Corning Toray). The device consisted of two layers: the upper layer for the control of the membrane valves and the lower layer for the flow of solutions. Molds for both layers were fabricated based on CAD designs using a computer controlled machining. The molds for the upper and lower layers were machined from aluminum and acrylic, respectively, using a 5-axis CNC milling machine (D200Z, Makino).
The channel heights were designed as follows: the upper control layer was 80~\textmu m, while the lower flow layer was 10~\textmu m in the capillary and outlet channels, {\color{blue}50~\textmu m in the observation chamber}, and 30~\textmu m in the other channels.
The channel widths around the flow focusing junction for droplet generation were designed as follows: the water inlet width was 300~\textmu m, the oil inlet was 200~\textmu m, and the outlet, where the droplets are generated and subsequently connected to the chamber, was 50~\textmu m (Fig.~\ref{fgr:fig2_verification}A).
The PDMS base and curing agent were mixed at a ratio of 5:1 for the upper layer and 20:1 for the lower layer, and the mixtures were degassed using a vacuum desiccator. 
The mixture for the lower layer was then spin-coated on the lower layer mold at 300 rpm for 30 s and then 800 rpm for 30 s. 
The mixture for the upper layer was poured on the upper-layer mold. 
Both molds were soft-baked at \SI{90}{\celsius} for 17 min. 
After the soft-bake, the upper-layer PDMS was immediately peeled off from the mold, and the inlet holes were punched with a biopsy punch with 1.5 mm diameter hole (BP-15F, Kai industries). 
The punched PDMS replica was then placed on the lower-layer mold for curing at {\color{blue}\SI{90}{\celsius}} for 2 h.
The cured PDMS was then peeled from the mold, and the inlet holes were punched using a biopsy punch with 1.0 mm diameter hole (BP-10F, Kai industries). 
Finally, the PDMS replica was bonded with a glass slide using plasma for 10 s using a plasma irradiation device (PC-400T, Strex).
The device was further heat-cured overnight to improve the bonding.

\subsubsection*{Sample preparation}

The DNA thermometer was purchased from Eurofins Genomics K.K. (Japan) with both ends modified by a fluorescent molecule (TAMRA) and a quencher molecule (Black Hole Quencher 2, BHQ2), respectively, and purified via HPLC. The DNA sequence of the thermometer is provided in {\color{blue}Table S6} in Supplementary Note~6, which was adopted from that reported in literature \cite{Goddard2000}. The DNA thermometer was diluted to a concentration of 100 {\textmu}M in a 100 mM phosphate buffer solution at pH 6.5 and stored at \SI{-20}{\celsius}.
In experiments, the DNA thermometer was further diluted to a final concentration of 2 {\textmu}M using a phosphate buffer solution containing 100 mM sodium ions at pH 6.5.
The enzymatic reaction system utilized $\beta$-glucosidase (BGH-101, Toyobo) and fluorescein-di-$\beta$-D-glucopyranoside (ab273893, Abcam). In experiments involving the enzymatic reaction system (Fig. \ref{fgr:fig3_constant_gradient} and Fig. \ref{fgr:fig4_change_gradient}), the reaction components were diluted in a 100 mM phosphate buffer solution at pH 6.5 with 100 mM sodium ions, achieving final concentrations of $5.0 \times 10^{-2}$ g/L for $\beta$-glucosidase and $5.0 \times 10^{-3}$ g/L for fluorescein-di-$\beta$-D-glucopyranoside.
The oil phase used for droplet formation consisted of fluorinated oil (HFE7500, Novec) with 4 w/w$\%$ of a fluorinated surfactant (FluoSurf, Emulseo).

\subsection*{Device setup and droplet generation}

The control layer of the microfluidic device was filled with double distilled water.
A polytetrafluoroethylene (PTFE) tubes (Chukoh Chemical Industries) were connected to the inlets of the control layer and linked to a pressure regulation system. Specifically, each tube was connected to a solenoid valve, which in turn was connected to a pressure regulator and then to an air compressor. The solenoid valve was actuated by a custom-built electric circuit to enable pneumatic control of the membrane valves.
Separately, PTFE tubes were attached to syringes containing the oil phase. The syringes were mounted on syringe pumps (NE1000, NEW ERA Pump Systems).
After the aqueous sample solutions for droplet encapsulation were preloaded into the tubing to avoid air bubbles, the PTFE tubes were connected to the inlets of the flow layer.
The microfluidic device was then placed on a stage incorporating two Peltier elements.

\subsection*{Thermal control and imaging conditions}
An electric circuit was built to regulate the temperature of the two Peltier elements (see Supplementary Note 7, {\color{blue}Fig. S9}). The platinum resistance thermometers were attached to the surface of the Peltier elements, and the applied voltage was controlled using a custom Raspberry-Pi based feedback system.
The distance between the two Peltier elements was set to 20 mm for the experiments shown in Fig. \ref{fgr:fig2_verification}B, Fig. \ref{fgr:fig3_constant_gradient}, and Fig. \ref{fgr:fig4_change_gradient}, and to 3 mm for the experiments shown in Fig. \ref{fgr:fig2_verification}C to establish isothermal conditions across the observation chamber.

The syringe pump was activated to generate droplets in the chamber. Once the chamber was filled with droplets, the control layer was pressurized to 200 kPa, and the syringe pump was stopped. Immediately after this operation, the PTFE tubing connected to the flow layer was clamped to immobilize the droplets in place.

In each experiment, the temperature setpoints of the two Peltier elements were configured as follows. For the experiments in Fig. \ref{fgr:fig2_verification}B, the temperatures were set to \SI{30}{\celsius} and \SI{60}{\celsius}, respectively, after droplet generation. For Fig. \ref{fgr:fig2_verification}C, both Peltier elements were initially set to \SI{60}{\celsius} after droplet generation. After 3 min, bright-field and fluorescence images were acquired. Then, the temperature was lowered in \SI{2}{\celsius} steps from \SI{60}{\celsius} to \SI{30}{\celsius} with images captured 3 min after each adjustment. 
For the stationary temperature gradient in Fig. \ref{fgr:fig3_constant_gradient}, the two Peltier elements were preheated at 
\SI{30}{\celsius} and \SI{60}{\celsius}, respectively, during droplet formation. Bright-field and fluorescence images were acquired every 5 min for 60 min after the droplet immobilization. Image acquisition at 50~min was not available due to a technical issue.
For Fig. \ref{fgr:fig4_change_gradient}, the initial setpoint of the temperatures was \SI{30}{\celsius} and \SI{65}{\celsius}, respectively. Bright-field and fluorescence images were acquired every 2.5 min for 77.5 min, starting from the initiation of heating. During this process, the temperature settings of the Peltier elements were alternated between \SI{30}{\celsius} and \SI{65}{\celsius} every 30 min.

\subsection*{Data acquisition}

Bright-field and fluorescence images were acquired using an inverted epifluorescence microscope (Eclipse Ti-S, Nikon) with an sCMOS camera (pco.panda, Excelitas PCO). The exposure times were 1 ms (bright-field), 1000 ms (TAMRA: ex 540/25 nm, em 605/55 nm), and 250 ms (fluorescein: ex 469/35 nm, em 525/39 nm). All image acquisition was automated using Micro-Manager software \cite{edelstein2010computer}.

To image the entire observation chamber, bright-field images from multiple fields of view were acquired by scanning the auto x-y stage (BIOS-106T, Sigma Koki) and then stitched into a single composite image using the grid/collection stitching plugin in ImageJ Fiji \cite{Fiji}. {\color{blue}For Figs.~3 and 4, droplets were detected from individual bright-field images before stitching, and the detected droplet positions were then integrated after removing redundant entries from overlapping regions between adjacent fields of view.} {\color{blue}For Fig. 2, droplets were detected from a single bright-field image. }
{\color{blue}Droplet detection was performed} by binarization followed by particle analysis (Analyze Particles function in Fiji) and their {\color{blue}centroid} positions were recorded. {\color{blue} 
Predefined size and circularity criteria, applied uniformly to all images within each experiment, were used to identify individual droplets.
The size range was set to single droplets ($200$--$600~\mathrm{pixels}^2$ for Fig.~2B, $50$--$500~\mathrm{pixels}^2$ for Figs.~2C and 4, and $50$--$750~\mathrm{pixels}^2$ for Fig.~3), 
and circularity thresholds of 0.70 for Figs.~\ref{fgr:fig2_verification} and \ref{fgr:fig4_change_gradient} and 0.60 for Fig.~\ref{fgr:fig3_constant_gradient} were applied.} 
{\color{blue}Droplets whose DNA-thermometer fluorescence intensities exceeded the range of the calibration curve in Fig.~2C were not assigned temperatures and were excluded as missing values.} 
{\color{blue}The lateral edges of the observation chamber were excluded prior to fluorescence intensity quantification. The excluded width was 15\% on each side for the DNA thermometer calibration in Fig.~2C and 5\% on each side for the experiments in Figs.~3 and 4 (see also Training and validation of Neural ODE model section).}

\subsection*{Training and validation of Neural ODE model}

For the training of Neural ODE, time-series trajectories {\color{blue}corresponding to 
droplets located within the outermost 5\% of the observation chamber on both lateral sides were excluded (Figs. S4 and S8). }
The trajectories were then divided into training, validation, and test sets with a ratio of 8:1:1.
The Neural ODE was implemented in PyTorch using \texttt{torchdiffeq}.
All computations were performed with reproducible random seeds (\texttt{SEED = 42}) and 
the options \texttt{torch.backends.cudnn.deterministic = True} and \texttt{benchmark = False}, which ensures deterministic behavior across CPU and GPU executions.
The reaction rate function $f_\theta(S(t), T(t))$ was parameterized by a neural network with sigmoid linear unit (SiLU) activation functions followed by a Softplus-shifted output layer (\texttt{Softplus}($\beta = 0.5$) + $\varepsilon$ with $\varepsilon = 10^{-3}$) to enforce strictly positive reaction rates. {\color{blue}For the stationary temperature-gradient experiment in Fig.~\ref{fgr:fig3_constant_gradient}, the neural network consisted of two hidden layers with 64 neurons per hidden layer. For the dynamic temperature-gradient experiment in Fig.~\ref{fgr:fig4_change_gradient}, the neural network consisted of three hidden layers with 128 neurons per hidden layer. 
}
Model parameters were optimized using the Adam optimizer with an initial learning rate $10^{-4}$ and a batch size of 64 for up to 400 epochs.  
A learning rate scheduler (\texttt{ReduceLROnPlateau}, factor = 0.5, patience = 5) was employed to automatically reduce the learning rate when the validation loss did not improve for five consecutive epochs.  
The training loss was defined by a weighted mean squared error of the form 
\begin{align*}
L = \frac{1}{N} \sum_{i=1}^{N} w_{b(i)} 
\left( y_i^{\mathrm{obs}} - y_i^{\mathrm{pred}} \right)^{2} \ \ \mathrm{with} \ \ 
w_b = 
\frac{\dfrac{1}{n_b}}{\dfrac{1}{B} \sum_{j=1}^{B} \dfrac{1}{n_j}},
\end{align*}
where $N$ is the total number of all time points and droplets, $y_i^{\mathrm{obs}}$ and $y_i^{\mathrm{pred}}$ denote the observed and predicted fluorescence intensities, respectively, 
$b(i)$ indicates the temperature bin of sample $i$, 
$n_b$ is the number of samples in bin $b$, 
and $B$ is the total number of bins. The weight $w_b$ is intended to compensate for imbalances in the number of samples across temperature bins. 
{\color{blue}
All computations were performed on Google Colaboratory using PyTorch 2.10.0 with CUDA 12.8 on an NVIDIA A100-SXM4-40GB GPU. The runtime was equipped with 12 logical CPU cores and 83.47 GB of system memory. The approximate wall-clock training time, measured from the start of parameter optimization to the end of training, was approximately 13 min for the stationary temperature-gradient dataset and approximately 23 min for the dynamic temperature-gradient dataset.
}

\section*{Author contributions}
\textbf{Mamoru Saita}: Methodology, Software, Formal Analysis, Investigation, Writing - Original Draft, Visualization.\\
\textbf{Yutaka Hori}: Conceptualization, Methodology, Formal Analysis, Writing - Review \& Editing, Supervision, Funding acquisition.

\subsection*{Data availability}
The codes and data that support the findings of this study are available {\color{blue}at 
\url{https://github.com/hori-group/neural-ode-temperature-kinetics}
}
\subsection*{Financial disclosure}
This work was supported in part by JSPS KAKENHI Grant Number JP18H01464, JP21H05889, JP23H00506, JP24K00911, JP26H00483, and JST-Mirai Program Grant Number JPMJMI22G6, JPMJMI25H2.

\subsection*{Conflict of interest}
The authors declare no conflict of interests.

\bibliographystyle{unsrt}
\bibliography{saita} 
\end{document}